\documentclass[preprint2]{aastex}
\usepackage{epsfig}
\usepackage{psfig}
\usepackage{graphicx}

\newcommand{\etal}{{\em et al.}}

\def\aj{{AJ}}

\def\annrev{{ARA\&A}}
\def\apj{{ApJ}}
\def\apjs{{ApJS}}

\def\lax{{$\mathrel{\hbox{\rlap{\hbox{\lower4pt\hbox{$\sim$}}}\hbox{$<$}}}$}}
\def\gax{{$\mathrel{\hbox{\rlap{\hbox{\lower4pt\hbox{$\sim$}}}\hbox{$>$}}}$}}
\def\simlt{\lower.5ex\hbox{$\; \buildrel < \over \sim \;$}}
\def\simgt{\lower.5ex\hbox{$\; \buildrel > \over \sim \;$}}

\def\mnras{{MNRAS}}

\def\percm2{cm$^{-2}$}

\slugcomment{To be submitted to {\it The Astrophysical Journal}}
\shorttitle{\textit{Chandra} Observations of 3C\,288}
\shortauthors{Lal et al.}

\begin{document}

\title{A \textit{Chandra} Observation of 3C\,288 -- Reheating the Cool Core
of a 3~keV Cluster from a Nuclear Outburst at z = 0.246}

\stepcounter{footnote}

\author{D. V. Lal$^{1}$; R. P. Kraft$^{1}$; W. R. Forman$^{1}$;
M. J. Hardcastle$^{2}$; C. Jones$^{1}$; P.~E.~J. Nulsen$^{1}$;
D.~A.~Evans$^{1,3}$; J. H. Croston$^{4}$; J. C. Lee$^{1}$}
\affil{$^{1}$Harvard-Smithsonian Center for Astrophysics, 60 Garden Street, Cambridge, MA 02138, USA}
\affil{$^{2}$University of Hertfordshire, School of Physics, Astronomy, and Mathematics, Hatfield AL 10 9AB, UK}
\affil{$^{3}$MIT Kavli Institute for Astrophysics and Space Research, 77 Massachusetts Avenue, Cambridge, MA 02139, USA}
\affil{$^{4}$University of Southampton, School of Physics and Astronomy, Southampton, SO17 1SJ U.K.}

\begin{abstract}

We present results from a 42 ks \textit{Chandra}/ACIS-S observation
of the transitional FR\,I/FR\,II radio galaxy 3C\,288 at $z$ = 0.246.  
We detect $\sim$3~keV gas extending to a radius of $\sim$0.5~Mpc
with a 0.5--2.0 keV luminosity of 6.6 $\times$ 10$^{43}$ ergs~s$^{-1}$,
implying that 3C\,288 lies at the center of a poor cluster.
We find multiple surface brightness discontinuities in the gas indicative of
either a shock driven by the inflation of the radio lobes or a recent
merger event.  
The temperature across the discontinuities is roughly constant
with no signature of a cool core, thus disfavoring
either the merger cold-front or sloshing scenarios.
We argue therefore that the discontinuities are shocks due to
the supersonic inflation of the radio lobes.
If they are shocks, the energy of the outburst is $\sim$10$^{60}$ ergs,
or roughly 30\% of the thermal energy of the gas within the radius of the shock,
assuming that the shocks are part of a front produced by a single outburst.
The cooling time of the gas is $\sim$10$^8$ yrs,
so that the energy deposited by the nuclear outburst could have reheated and
efficiently disrupted a cool core.

\end{abstract}

\keywords{galaxies: individual (3C\,288) - X-rays: galaxies: clusters -
galaxies: IGM - hydrodynamics - galaxies: jets}


\section{Introduction}
\label{sec:intro}

\textit{Chandra} has revolutionized our knowledge of the energetic
processes involved in the jets, the lobes and the nuclei of radio galaxies.
The X-ray images of the hot atmospheres in galaxies show a wealth
of structures associated with central radio sources, including cavities,
metal enriched plumes, filaments, and shock fronts 
\citep{peterson06,mcnamara07,mcnamara09}.
High-resolution spectroscopic observations from \textit{Chandra} and
\textit{XMM-Newton} have conclusively ruled out simple, steady cooling
flow models and this has been one of the significant discovery
\citep{peterson06,david06}.
Since the cooling time of gas in many cluster cores is much less
than the Hubble time, energy must be occasionally or continuously
supplied to cluster cores to prevent the formation of cooling flows.
A primary candidate for the suppression of cluster cooling flows is feedback
between the central supermassive black hole (SMBH) of active galaxy and
the cluster gas.

Studies of the X-ray gas environments of FR\,I and FR\,II
sources with \textit{Chandra} have led to important new constraints
on how jets propagate through their ambient media, and
how radio lobes interact with and transfer energy to their
large-scale gas environments \citep[e.g.,][]{mjh02,sambruna04,kraft06}.
However, `high-excitation'
FR\,II sources tend to lie in low gas mass atmospheres at least at low
redshifts \citep{ellingson91a,ellingson91b,ellingson91c,harvanek01,harvanek02,isobe05,kraft07b}, unless the mechanical power
of the jet is unusually high \citep[e.g., Cygnus~A][]{smith02}.
This strongly suggests that the jet power and the interaction
between jets and the hot gas in their vicinity play pivotal roles in
governing the overall morphology (FR\,I vs. FR\,II) of the radio source.

The radio source 3C\,288 is identified with an elliptical galaxy with
$m_v$ = 18.3 \citep[$M_v$ = $-$22.08,][]{goodson79}.
It is the archetypal example of a
transitional FR\,I/FR\,II (`jetted double') radio galaxy
\citep{fr74}.
In optical imaging, there are fainter galaxies in the field which are
presumed to be members of a cluster that is dominated by 3C\,288
\citep{wyndham66}.
Dominant cluster members can produce asymmetric ``wide-angle-tail''
(WAT) radio structures \citep{burns86}, despite their presumed
low peculiar velocities, but 3C\,288 is not a clear example
of WAT morphology.
Its monochromatic power at 1.5 GHz is
2.5 $\times$ 10$^{33}$ ergs~s$^{-1}$~Hz$^{-1}$ and its
integrated spectral index $\alpha^{5.0}_{0.75}$ between 0.75 GHz and
5.0~GHz is $-$0.97 \citep[][$s_\nu \propto \nu^\alpha$]{laing80}.
Its radio luminosity would place it firmly in the high-power
(i.e., FR\,II) regime, and yet its observational characteristics,
in particular its edge-darkened radio morphology,
are strikingly different from those of the canonical `classical double'
radio galaxies.
Although, the edge darkening of the radio structure, and
its spectral distribution, are reminiscent of a (distorted)
radio ``trail'', no other properties of 3C\,288 suggest that
it belongs to this morphological class.
The structure of 3C\,288 is more asymmetric than those of normal
double sources of its size and radio power \citep{bridle89}.
VLA observations reveal a jet and a counterjet
near the radio core, and faint ``wings'' of emission connected
to the elongated lobes \citep{bridle89}.
The small radio size, unusual morphology, and polarization
asymmetries of 3C\,288 raise the possibility that it is
interacting strongly with the ambient gas \citep{bridle89}.
Thus, 3C\,288 is an ideal candidate for study with \textit{Chandra} to
better understand
the role that AGN heating may play in the formation of radio structure
in the hot phase of the intergalactic medium (IGM).

This paper is organized as follows:
Section~2 contains a summary of the observational details.
Results of the data analysis are presented in Section~3 and
we discuss their implications in Section~4.
Section~5 contains a brief summary and conclusions.
We assume \textit{WMAP} cosmology throughout this paper \citep{spergel07}.
The observed redshift ($z$ = 0.246) of the host galaxy of 3C\,288
corresponds to a luminosity distance of 1192.9 Mpc, and
1$^{\prime\prime}$ = 3.725 kpc.  All coordinates are J2000.
The elemental abundances are relative to the Solar value
tabulated by \citet{anders89}.  Absorption by gas in
our galaxy \citep[$N_{\rm H}$ = 9 $\times$ 10$^{19}$ cm$^{-2}$,][]{dickey90} is
included in all our spectral fits.

\section{Observations} \label{observation}

The radio galaxy 3C\,288 was observed on 2008-04-13 (OBSID 9275; P.I. D.A. Evans) with
\textit{Chandra}/ACIS-S in VFAINT mode for $\sim$42.0 ks.
We made light curves for each CCD in the 0.5 keV to 10.0 keV band in order to
search for background flares and intervals where the background rate was high;
none were present, leaving 39647.8 s of good data.
We performed the usual filtering by grade, excluded bad/hot pixels and columns,
removed cosmic ray ``afterglows'', and applied the VF mode filtering using
tools built in CIAO (http://cxc.harvard.edu/ciao).
Images were generated after subtracting background and correcting for exposure
(which included all the effects mentioned above).

We use archival VLA observations of 3C\,288 at 4.885~GHz \citep{bridle89}.
The map is taken from the online 3CRR Atlas\footnote{ATLAS catalog:
Radio images and other data for the nearest 85 DRAGNs
(radio galaxies and related objects) in the so-called ``3CRR"
sample of \citet{laing83}. Available at
{\tt http://www.jb.man.ac.uk/atlas/index.html} .}
which provides well-calibrated, well-sampled images.

\section{Data Analysis} \label{data_a}

\begin{figure}
\begin{center}
\begin{tabular}{l}
\includegraphics[width=7.24cm]{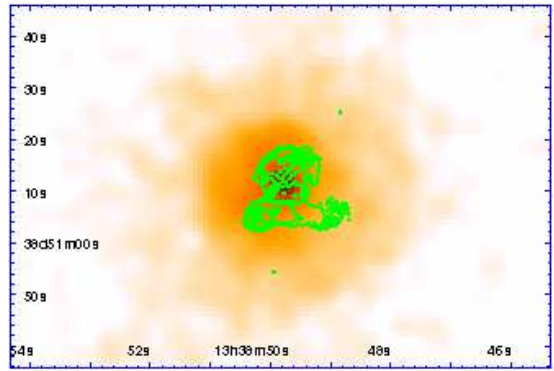}
\end{tabular}
\end{center}
\caption{Gaussian-smoothed (\textrm{radius$_{\rm ~FWHM}$} = 2$^{\prime\prime}$),
exposure corrected, background subtracted
\textit{Chandra}/ACIS-S image of 3C\,288 in the 0.5--2.5 keV band.
All point sources, other than the active nucleus of 3C\,288, have been removed.
We see diffuse thermal emission with a temperature of $\sim$3 keV extending to
$\sim$510 kpc ($\sim$1.73$^{\prime}$).
Contours from the 4.9~GHz radio map of 3C\,288 are overlaid and 10
contour levels are placed linearly between 0.4 and 12.0 mJy~beam$^{-1}$.
The radio source is small, 294 kpc (1.3$^{\prime}$ across) and the radio core
is conincident with the optical host galaxy.
A peak of X-ray emission lies $\sim$11.2 kpc
($\sim$3.0$^{\prime\prime}$) east of the optical host galaxy.}
\label{rad_on_xray}
\end{figure}

A Gaussian-smoothed ($\textrm{radius$_{\rm ~FWHM}$} = 2^{\prime\prime}$) \textit{Chandra}/ACIS-S
image of 3C\,288 in the 0.23--5.00~keV band with radio contours overlaid is
shown in Figure~\ref{rad_on_xray}.
All point sources other than the 3C\,288 nucleus have been removed.
We detect diffuse thermal emission from the cluster ICM with a temperature of $\sim$3~keV extending to
$\sim$390 kpc ($\sim$104$^{\prime\prime}$).
Diffuse emission from the source fills only a fraction ($\sim$3\%) of the area in the S3-chip,
so local background was extracted for all spectral analyses,
from the source-free region on the S3 chip.
The radio source is small,
$\sim$ 67.1 kpc (= 18$^{\prime\prime}$ across), and the X-ray
bright gas core lies $\simeq$ 11.2~kpc
($\sim$3$^{\prime\prime}$) north of the radio core; this
offset is perhaps associated with the non-hydrostatic motion of the gas.
The X-ray isophotes within 112~kpc (0.5$^\prime$) of the nucleus are circular.
For simplicity we assume spherical symmetry in our analysis below.
On larger scales, however, the isophotes show an extension to the south-east.

\begin{figure}
\begin{center}
\begin{tabular}{l}
\includegraphics[width=7.24cm]{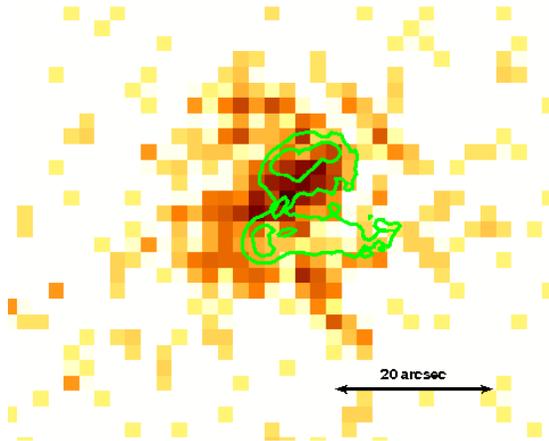}
\end{tabular}
\end{center}
\caption{\textit{Chandra}/ACIS-S image of 3C\,288 in the 0.23--0.50 keV
soft X-ray band with 5~GHz radio contours (0.6$^{\prime\prime}$
resolution) overlaid. This image shows the soft X-ray excess,
which is not commonly observed in clusters of galaxies and
is uniformly detected across the whole 42~ks observation.}
\label{soft_excess}
\end{figure}

Interestingly, we detect significant X-ray emission below 0.50 keV.  Such emission
is not commonly observed in clusters of galaxies with \textit{Chandra}, and
was initially suggestive of
inverse Compton scattering of CMB photons from a large population of cosmic ray electrons in the cluster
core, similar to that claimed for the Coma cluster
\citep{sarazin98,finoguenov03,erlund07}.
This emission lies predominantly at the cluster core.
Additionally, it appears that there is a cavity in the gas associated with
the southern radio lobe and that the 0.23--0.50 keV emission is roughly
aligned with the jet axis.
However there is no obvious direct correspondence
between this soft X-ray emission and radio features as seen in Hydra~A
\citep{nulsen2005}.
An image of the cluster in the 0.23--0.50 keV
band with radio contours overlaid is shown in Figure~\ref{soft_excess}.
We created surface brightness profiles of the emission in the
0.3--0.5 keV (soft) and 0.5--1.5 keV (hard) bands,
subtracted an appropriate background,
and fitted a line to the ratio of soft and hard X-ray emission as a
function of distance from the center.  The slope of this line is
consistent with zero within the uncertainties and we confirm that,
within the statistical uncertainties,
there is no difference in the spatial distribution of this soft X-ray emission
relative to the hard X-ray emission.
Spectral analysis of the \textit{Chandra} data, combined with spectral analysis of archival
\textit{ROSAT PSPC} data, confirm that this emission is simply the
bremsstrahlung continuum emission from a $\sim$3 keV plasma.  
It is only visible in the 3C\,288 cluster because of the combination of gas
temperature and unusually low Galactic column.

\subsection{Compact Components}

Contours from the 4.9~GHz radio map of 3C\,288 are shown in
Figure~\ref{rad_on_xray} overlaid on the background-subtracted,
exposure-corrected, Gaussian-smoothed \textit{Chandra}~ACIS-S image.
The core is the brightest, most compact radio feature \citep{bridle89}
and coincides with the optical position of the nucleus 
\citep{goodson79}.
It has a flux density, $S_{\rm 4.9\,GHz}$, of 12.68 $\pm$0.07 mJy
(1 mJy = 10$^{-26}$ ergs~s$^{-1}$~cm$^{-2}$~Hz$^{-1}$).
The northern and southern hotspots are clearly detected in
4.9~GHz and 1.4~GHz radio maps and have flux densities of
6.88 $\pm$0.09 mJy and 5.84 $\pm$0.11 mJy, respectively at 4.9\,GHz.
The radio core has a spectral index of $\alpha^{\rm 5.0~GHz}_{\rm 1.4~GHz}$
= $-$0.76 $\pm$0.02 ($s_\nu \propto \nu^\alpha$,
where $S_\nu$ is the flux density at frequency $\nu$; and the error~bars are 1$\sigma$ confidence).
The northern and southern hotspots 
have spectral indices of $\alpha^{\rm 5.0~GHz}_{\rm 1.4~GHz}$
= $-$0.76 $\pm$0.04 and $-$0.87 $\pm$0.06, respectively.
We do not detect an X-ray point source (see Figure~\ref{rad_on_xray})
at the location of the radio core.
Only eight net counts above background
were extracted in a 0.5$^{\prime\prime}$ radius
circle at the location of the core with no point source evident above
the level of the diffuse emission from the gas.
The 3$\sigma$ upper limit to the 1~keV (rest frame) flux
density of the core is $\sim$0.31~nJy assuming a photon index of 2 and Galactic
absorption.  Assuming that this upper limit represents nonthermal X-rays from the AGN,
the upper limit to the radio to X-ray power-law index is
$\alpha^{\rm X-ray}_{\rm radio}$ $\simeq$ $-$0.99 $\pm$0.02.
The corresponding core radio luminosity at 178 MHz assuming a flat spectrum
and the unabsorbed core X-ray luminosity at 1~keV are consistent with the
expectation from the correlation between these two quantities shown in
\citet{mjh09}.

No X-ray emission is detected from the jet or the compact hot spots of
the two lobes.
With no point source evident and zero counts above the diffuse emission
from the gas,
we use the 3$\sigma$ counts of seven, by
measuring the off-source background level and then use Poisson
statistics to find the number of counts corresponding to a 3-sigma
Gaussian probability, for 
the northern and the southern hotspots.
The 3$\sigma$ upper limit to the 1~keV (rest frame) flux density of both,
the northern and southern hotspots is $\sim$0.29~nJy
assuming a photon index of 2 and Galactic absorption.
We deduce upper limits to the X-ray--radio power-law indices for
the northern and southern hotspots of
$\alpha^{\rm X-ray}_{\rm radio}$ $\gtrsim$ $-$0.96 $\pm$0.02
and $\gtrsim$ $-$0.95 $\pm$0.02, respectively (again
the error~bars are at 1$\sigma$ confidence).
If we use these spectral indices, the expected flux densities
in the optical and infrared bands,
$S_{5000\,{\buildrel _{\circ} \over {\mathrm{A}}}}$ and $S_{7\,\mu}$ are
0.09 and 1.15 $\mu$Jy for the northern hotspot, and 0.09 and 1.07 $\mu$Jy
for the southern hotspot, respectively.
These are undetectable with the current generation of optical and infrared observatories.
There is no non-thermal emission detected from the nucleus, lobes, or jets from this radio
galaxy, and the upper limits are consistent with detections of such emission in
much closer radio galaxies.

\subsection{Large-Scale Diffuse X-ray Emission}

\begin{figure}[ht]
\begin{center}
\begin{tabular}{l}
\includegraphics[width=5.24cm,angle=-90]{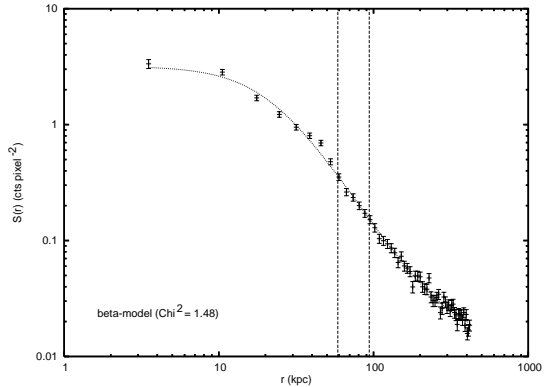}
\end{tabular}
\end{center}
\caption{Azimuthally averaged radial surface brightness profile of the X-ray
emission from the gas in the energy band from 0.5~keV to 2.5~keV.
The best-fit isothermal $\beta$-model profile has been overlaid.
The two vertical lines show the location of prominent, sharp edges
in X-ray brightness at distances from the nucleus of 53.3 kpc
(14.2$^{\prime\prime}$) and 91.1 kpc (24.4$^{\prime\prime}$)
along the eastern and the south-western directions, respectively.}
\label{sdf}
\end{figure}

The extended, diffuse X-ray emission seen in Figure~\ref{rad_on_xray}
is attributed to emission from the hot gas of a cluster atmosphere.
We derived a global temperature and metallicity for 3C\,288 within a
1$^\prime$ radius circular region covering the majority of the
cluster emission.  The spectrum was extracted using the
{\sc CIAO}\footnote{All spectral extraction and spectral analsis were performed using {\sc CIAO} v4.1, the {\sc CALDB} v4, and {\sc XSPEC} v12.5.1.}
specextract tool, binned to 10 cts per bin and fitted in the 0.5--5.0 keV
range using an absorbed {\sc APEC} model within the {\sc XSPEC} package
\citep{arnaud96}.
The neutral hydrogen column density was fixed at the Galactic foreground
value of $N_{\rm H}$ = 9.0 $\times$ 10$^{19}$ cm$^{-2}$.
The best fitting values for temperature and abundance are
$k_BT$ = 2.94$^{-0.18}_{+0.19}$~keV and $Z$ = 0.73$^{+0.16}_{-0.19}$,
respectively,
where the errors are 90\% confidence limits.
If we allow the neutral hydrogen column density to be a free parameter,
the best fitting values for temperature and abundance do not
change appreciably, and the changes are smaller that 1$\sigma$ uncertainties.
The unabsorbed X-ray luminosity from the best fitting model in the energy range
0.5--5.0 keV is 1.11 $\pm0.03$ $\times$ 10$^{44}$ ergs~s$^{-1}$ within
the $r$ = 223.5 kpc (1$^{\prime}$) circular aperture.
Additionally, the unabsorbed X-ray luminosity in the energy range
0.5--2.4 keV is 7.60 $\pm0.18$ $\times$ 10$^{43}$ ergs~s$^{-1}$ and the result is consistent
with the expectations for a 3~keV cluster \citep{maxim98}.
The azimuthally averaged radial surface brightness profile of the X-ray
emission from the gas is shown in Figure~\ref{sdf}.
The best-fitting isothermal $\beta$-model profile has been overlaid.  We find
$\beta$ = 0.52 $\pm$0.02 and a core radius $r_0$ = 11.94$^{\prime\prime}$
$\pm$0.93$^{\prime\prime}$ from fitting the surface brightness profile between
3$^{\prime\prime}$ and 200$^{\prime\prime}$ from the nucleus.

\begin{figure}[ht]
\begin{center}
\begin{tabular}{l}
\includegraphics[width=7.24cm]{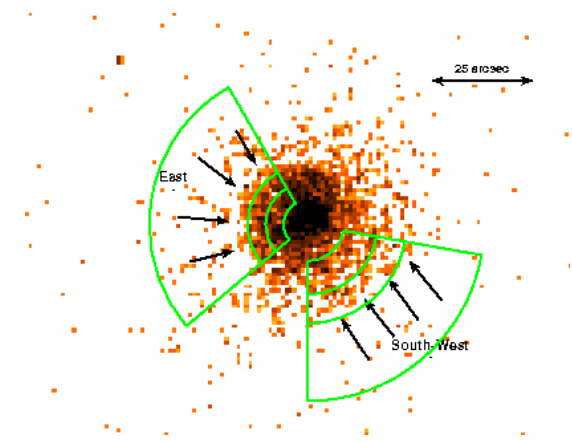}
\end{tabular}
\end{center}
\caption{Gaussian-smoothed (\textrm{radius$_{\rm ~FWHM}$} = 2$^{\prime\prime}$) \textit{Chandra}/ACIS-S
image in the 0.23--2.50 keV band.  The image has 2$^{\prime\prime}$ bins.
The black sets of arrows denote the positions of two surface brightness
discontinuities. The image also show regions where
surface brightness profiles were fitted, and where spectra were extracted
for temperature measurements.  A large 1$^\prime$ region centered on the
radio galaxy was used for global temperature and abundance measurements.
The surface brightness profiles shown in Figures~5 and 6 are taken from the two
sectors.}
\label{sbd}
\end{figure}

We find two breaks in the surface brightness distribution, one 14.2$^{\prime\prime}$ east of the
nucleus and another 24.4$^{\prime\prime}$ to the southwest, as shown by the black arrows in Figure~\ref{sbd}.
The presence of these breaks in the surface brightness distribution implies a sharp change in the gas density or temperature
of the gas across the discontinuities.  \textit{Chandra} has observed a large number of similar features
in other clusters, such as Abell\,1795 \citep{maxim01}, Abell\,3667 \citep{vikhlinin01a}, 
M\,87 \citep{forman2005,forman2007}, Hydra\,A \citep{nulsen2005}, MS\,0735.6$+$7421 \citep{mcnamara05},
and Abell\,1201 \citep{owers09}, and they are generally 
attributed to three phenomena:  merger cold fronts, sloshing
of cluster cores due to non-hydrostatic motions of the gas, and shocks due to nuclear outbursts.
Most merger cold fronts are offset from the center (e.g. Abell\,3667) and are the result of the
infall of a massive subclump into the cluster.  Unless we are viewing such a merger head-on, the
morphology of these features in the 3C\,288 cluster gas is very different than what we observe in Abell~3667 \citep{vikhlinin01a}.
Additionally, it would be surprising, if we are witnessing a major merger from such a viewing angle, that
the cluster lies on the $L_x$--$T$ relation.  Thus, we consider this possibility to be unlikely.  
Following the analysis of \citet{maxim07},
below we determine the temperature and pressure across the discontinuity to evaluate
which of the other two scenarios, sloshing or supersonic inflation
of radio lobes, is more plausible.

\begin{figure*}[ht]
\begin{center}
\begin{tabular}{l}
\includegraphics[width=16.0cm]{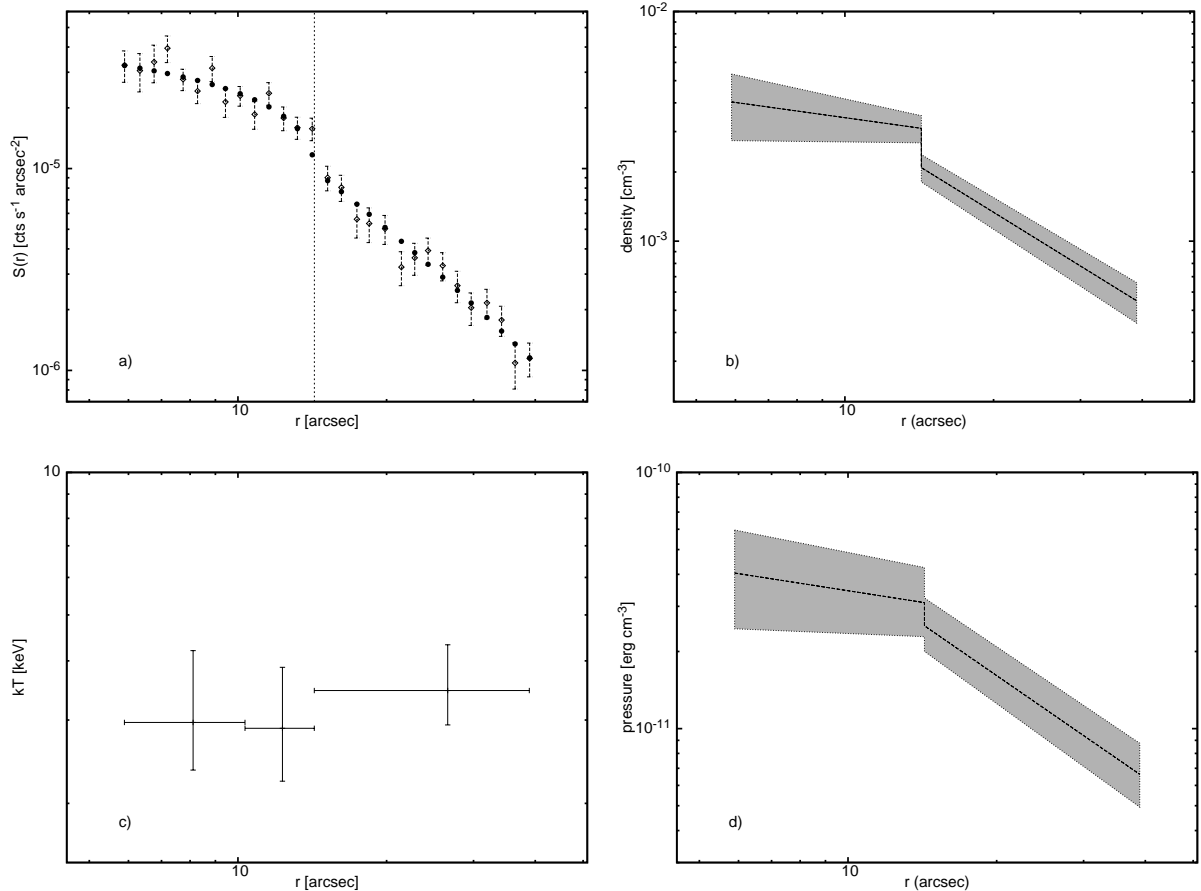}
\end{tabular}
\end{center}
\caption{Shock in 3C\,288 showing the eastern edge
(sector between P.A. = 30$^\circ$ and P.A. = 130$^\circ$).
Panel (a) shows X-ray surface brightness profile.  The filled
circles are the model values that correspond to the best-fit gas
density model shown in panel (b).  Panel (c) shows the temperature profile.
Panel (d) shows pressure profile obtained from the temperatures on either
side of the edge and density profile.
Error bars are 90\%;
vertical dashed line in panel (a) show the position of the density jump.}
\label{de_proj_e}
\end{figure*}

\begin{figure*}[ht]
\begin{center}
\begin{tabular}{l}
\includegraphics[width=16.0cm]{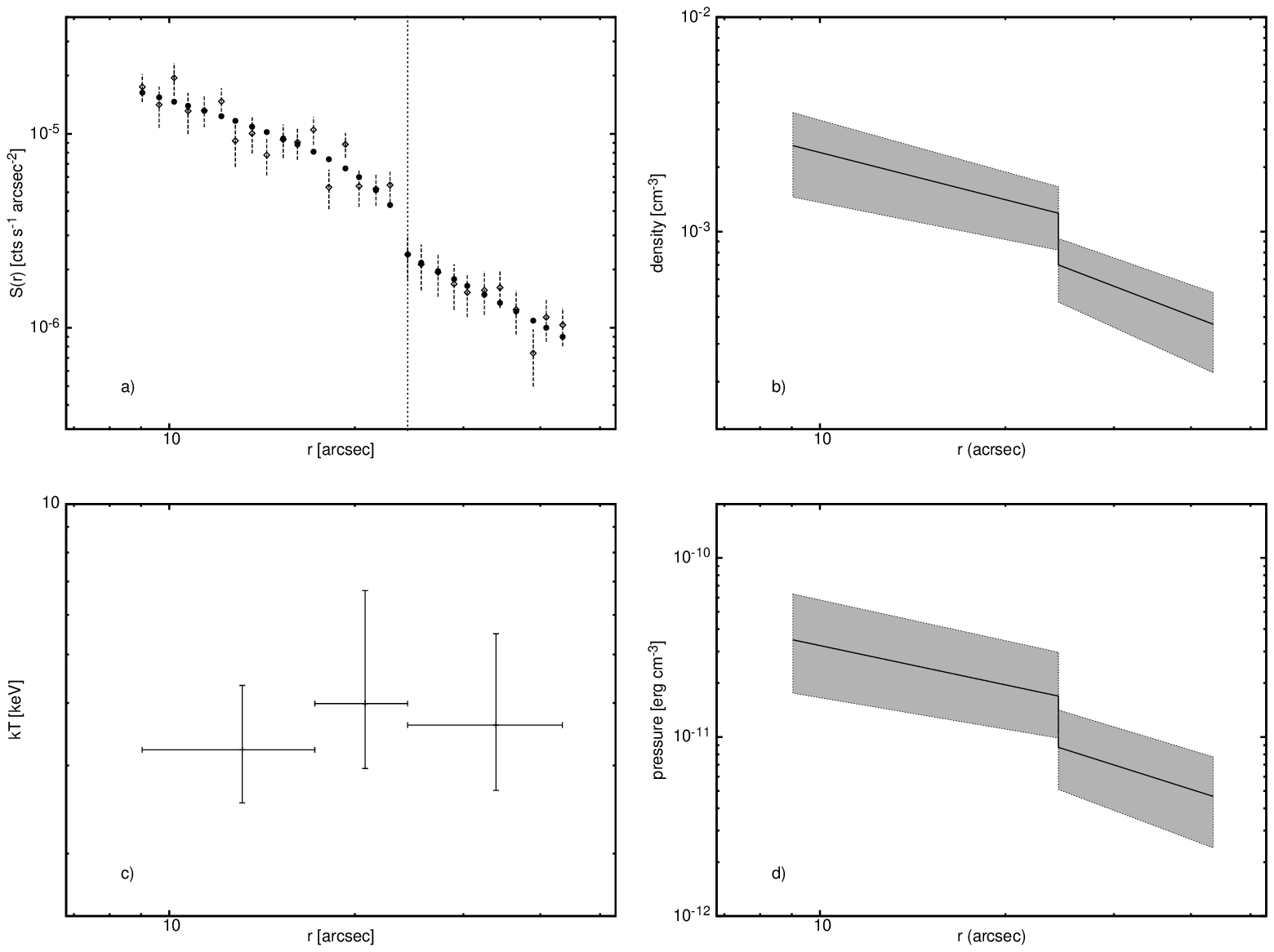}
\end{tabular}
\end{center}
\caption{X-ray properties in the south-western edge
(sector between P.A. = 180$^\circ$ and P.A. = 260$^\circ$).
Panel (a) shows X-ray surface brightness profile.  The filled
circles are the model values that correspond to the best-fit gas
density model shown in panel (b).  Panel (c) shows the temperature profile.
Panel (d) shows pressure profile obtained from the temperatures on either
side of the edge and density profile.
Error bars are 90\%;
vertical dashed line in panel (a) show the position of the density jump.}
\label{de_proj_sw}
\end{figure*}

We fitted absorbed {\sc APEC} models to three annuli in two sectors (shown in
Figure~\ref{sbd}) centered on 3C\,288.
The vertex of the annuli was fixed at the nucleus,
but the binning of the annuli was adjusted, so that the radius of curvature
of the second and third annular bins in each fit matches that of the associated
discontinuity (i.e. we created the bins so that two annuli were interior and
one annulus was exterior to the discontinuity).
The goal of this spectral fitting was to determine whether the gas temperature
interior to the discontinuities was hotter or cooler than the exterior gas temperature.
Only the temperatures and normalizations were free parameters in these fits.
The elemental abundance was frozen at the best-fit value determined in the
global fits ($Z$ = 0.73).
Plots of the temperature profiles for the two different sectors,
one between position angle (P.A.) = 30$^\circ$ and P.A. = 130$^\circ$
and another between P.A. = 180$^\circ$ and P.A. = 260$^\circ$,
as a function of radius from the phase-center (position of the host galaxy)
are shown in Figures~\ref{de_proj_e} and~\ref{de_proj_sw}, panel (c).
We find no significant jump in the projected temperature across either
discontinuity within 90\% error uncertainties of $\Delta T/T\sim$37\%.

We deprojected the surface brightness profiles (using the best-fit
gas temperature derived above)
to determine the density and pressure as a function of distance from
the nucleus.
Figure~\ref{sbd} shows the ACIS-S image of the central region, with prominent,
sharp edges in X-ray brightness at distances from the nucleus of 53.3 kpc
(14.2$^{\prime\prime}$) and 91.1 kpc (24.4$^{\prime\prime}$)
along the eastern and the south-western directions, respectively.
We model each of these surface brightness discontinuities with
a broken power-law density model. 
We fitted the surface brightness across the discontinuity in the
sectors shown in Figure~\ref{sbd} in the energy range 0.5--5.0 keV.
The deprojected density and pressure profiles between
30$^\circ$ and 130$^\circ$ (eastern direction),
and between 180$^\circ$ and 260$^\circ$ (south-western direction)
as a function of distance from the cluster center are
shown in Figures~\ref{de_proj_e} and~\ref{de_proj_sw}, panels (b) and (d), respectively.
Both brightness
profiles have a characteristic shape corresponding to a projection of an
abrupt, spherical (within a certain sector) jump in the gas density.
Best-fit radial density models of such a shape are shown in panel (b),
and their projections are overlaid on the data in panel (a) of
Figures~\ref{de_proj_e} and~\ref{de_proj_sw}.
From the amplitude of the best fitting surface brightness model,
we derived a density jump of 1.48$^{+0.28}_{-0.24}$ for the
eastern discontinuity and 1.75$^{+0.68}_{-0.39}$ for the south-western
discontinuity.  The confidence ranges for the density jumps were
computed from the extremes of the 90\% confidence ranges for
the best fitting surface brightness model.
As confirmation, the density profile that is derived from the isothermal, azimuthally symmetric
surface brightness profile is broadly consistent with the deprojections of the
broken power-law models.

To minimize projection effects, we would like to measure
the deprojected temperature profile across the jumps; however,
the limited number of photons simply does not permit this,
although the lack of temperature variation suggests that projection effects
are probably not large.
Instead, we used the extracted spectra from two large regions for each front for the 
subsequent analysis;
one for the bright side of the front (inside or post-shock) and one for the faint side
of the front (outside or pre-shock).  For the east jump, we measured a temperature of
2.88$^{+1.00}_{-0.66}$~keV inside the jump and 3.46$^{+0.86}_{-0.53}$~keV
outside the jump,
and for the south-western jump, we measured a temperature of
3.99$^{+2.72}_{-1.03}$~keV inside the jump and 3.61$^{+1.92}_{-0.94}$~keV
outside the jump.
The temperature ratios for the two discontinuities are 0.83$^{+0.36}_{-0.23}$
and 1.11$^{+0.96}_{-0.41}$ for the east jump and the south-western jump,
respectively.  These uncertainties are at 90\% confidence.

\section{Interpretation}

\subsection{Are the Discontinuities Shock Fronts or `Sloshing'?}

There are at least two possible explanations for the surface brightness discontinuities:
they could be shocks in the gas due to the supersonic inflation of the radio lobes,
or they could be contact discontinuities created by non-hydrostatic motions of the
gas core due to a recent merger (i.e. `sloshing').
Unfortunately the quality of the data is not sufficient for us to make a definitive statement
about whether these discontinuities are shocks or sloshing.  
The temperature of the cluster is sufficiently hot ($\sim$3 keV) that the cluster temperature is
not well constrained with less than several thousand counts per spectrum.  
We favor the shock model, as described below, but arguments can be made for either scenario.  We
describe some of the implications of both models.  Note that our conclusions about the
overall energetics of the radio galaxy are not significantly different under the two scenarios.
More explicitly, the estimate energy of the outburst is only tens of percent larger
for the shock scenario than the subsonic inflation scenario.

\subsection{Supersonic Inflation of the Radio Lobes}

Several lines of evidence suggest that the discontinuities are shocks,
including the relative symmetry of the discontinuities
around the nucleus which are not typically seen in sloshing cores, 
the lack of cool gas in the cluster core, the lack of any evidence of a recent
merger in the \textit{HST} image of the host galaxy \citep{dekoff96},
the lack of any evidence of merging on larger scales, and
the transformational morphology of the radio source from FR\,II to FR\,I suggesting
that it is strongly interacting with its environment.
If these features are shocks, we can estimate the expansion velocity of the radio lobes, total power of the
outburst, etc. to constrain the impact the effect of this outburst would have on the surrounding gas.  These conclusions
could be definitively confirmed (or refuted) with a deeper \textit{Chandra} or \textit{XMM-Newton} observation.

We use the example of M87 to provide a nearby analogy to interpret the features seen
in the 3C\,288 cluster gas.
M\,87 is the dominant central galaxy in the Virgo cluster
showing remarkable morphological details in both the X-ray and the
radio \citep{owen2000,hines1989,forman2005}.
Radio observations show evidence of two distinct nuclear outbursts, and
X-ray observations show at least two sets of surface brightness discontinuities in the
gas indicative of shocks \citep{forman2007}.
3C\,288 shows a signature of a shock, two regions of enhanced pressure,
at a radius of 53.3 kpc on the east and 91.1 kpc on the south-west,
similar to M\,87.

If these features are indeed shocks, the post-shock gas temperature and
density will be higher than the pre-shock values in a narrow region
behind the shock.
With the existing data, we cannot determine the
gas temperature profiles across the discontinuities with sufficient accuracy to
confirm that they are shocks.  Panel (c) in Figures~\ref{de_proj_e} and~\ref{de_proj_sw}
shows the gas temperature profiles across the edges.
For a shock discontinuity, the Rankine-Hugoniot jump conditions directly
relate the gas density jump, r $\simeq \rho_{\rm shock}/\rho_0$, and the
temperature jump, t $\simeq T_{\rm shock}/T_0$, where subscripts `0' and `shock'
denote quantities before and after the shock.  The
Mach number of the shock, $M$ $\equiv$ $v/c_s$, where $c_s$ is the
velocity of the sound in the pre-shock and $v$ is the velocity of the gas
with respect to the plane of the shock (e.g. Landau \& Lifshitz 1959).
Using the above density jumps, the Mach number is $M$ $\equiv$
1.33$^{+0.15}_{-0.12}$ and 1.53$^{+0.37}_{-0.21}$, respectively
for the east and south-west surface brightness discontinuities.
From these density jumps, we predict
temperature jumps $T_{\rm shock}/T_0$ $\equiv$
1.32$^{+0.38}_{-0.12}$ and 1.52$^{+1.16}_{-0.66}$,
respectively for the east and south-west wedges,
which are consistent, within error~bars, with our measured temperatures.
The pressure is discontinuous at the shock front as shown in
Figures~\ref{de_proj_e} and~\ref{de_proj_sw}, panel (d) for the east
and southwest shocks, respectively.

\subsubsection{Merging and `Sloshing'}

`Sloshing' of the dense cluster gas core is the term used to describe
the non-hydrostatic gas motions induced by a recent merger.
When a cluster undergoes a perturbation to its gravitational potential caused
by another infalling group or cluster, the gas core lags behind the cluster
the potential minimum, as they both move toward the perturbing object.
As the gas core falls back onto the potential minimum,
it overshoots it and begins to oscillate.
With each oscillation, the gas core is moving against its own trailing gas,
producing an ``edge" in the X-ray brightness which expands out from the cluster.
This sequence of events is described in more detail in \citet{AM06} and \citet{maxim01}.
The continued oscillation of the core gas about the potential minimum produces
a succession of radially propagating cold fronts, manifested as concentric
edges in the surface brightness distribution.  These fronts may form a spiral
structure when the sloshing direction is near the plane of the sky and the
merger has a non-zero angular momentum \citep{AM06}. 

The observational features of a cold front are a large temperature jump across the discontinuity
(the colder gas is closer to the nucleus) with no discontinuity in pressure.
The gas density, and therefore surface brightness, also generally forms a spiral pattern.
If we assume that the pressure is
continuous across the discontinuity, the temperature jumps
should be by factors of 0.67$^{+0.13}_{-0.11}$ and 0.57$^{+0.22}_{-0.13}$ respectively,
based on the derived density jumps for the eastern and the south-western surface brightness discontinuity.
These jumps are well within the uncertainties shown in Figures~\ref{de_proj_e} and~\ref{de_proj_sw}.
If we assume that we are viewing an advanced stage of gas sloshing,
the two discontinuities possibly form a clockwise spiral pattern around the core of the cluster
because the south-western discontinuity is father than the
eastern discontinuity from the core.
In much better exposed Chandra observations of `sloshing', however, the discontinuity is most prominent
in one small sector, not equally visible in two nearly opposite directions \citep{johnson10}

To our knowledge, there has been no comprehensive study of
individual galaxies of this cluster or measurement of their velocity dispersion.
The nearest galaxy to the cluster center detected in the Sloan Survey, SDSS~J133850.88$+$385216.0 ($z$ = 0.2442),
at a projected distance of $\sim$0.3~Mpc
(1.131$^{\prime}$) \citep[SDSS DR7;][]{Abazajian09}, has no gas
associated with it in the \textit{Chandra} image.
Additionally, there is no clear pattern of spiral structure in
the core, only an offset in the positions of the discontinuities.
However, the relatively low quality of the data makes a quantitative statement
impossible.  We conclude that the only compelling argument to support the sloshing hypothesis is the vague similarity
of the discontinuities seen in 3C 288 with other `sloshing' systems, and as described above there are
several lines of evidence to suggest they are shocks.  A much deeper Chandra observation and/or
a systematic study of the member galaxies of this cluster would provide a definitive answer.

\subsection{Nuclear Outburst}

If the surface brightness discontinuities in 3C\,288 are shocks,
we can estimate (i) the total energy and the age of the outburst,
(ii) the amount of mass accreted by the central SMBH, and
(iii) make a strong statement about the effects of outbursts on suppression
of the formation of large amounts of cool gas.
The total thermal energy of the gas within the core radius of the beta-model
is $\sim$2.0 $\times$ 10$^{60}$ ergs.  
This approach underestimates
the true energy because the shock front extends considerably
further to the east and southwest, implying a faster, stronger shock,
encompassing a greater volume in those directions.
However the value of 2.0 $\times$ 10$^{60}$ ergs provides a conservative estimate of the thermal
energy of the gas.
We estimate the mechanical energy of the outburst by two methods.  
First, assuming that the bubbles inflated adiabatically,
the total bubble enthalpy is $\sim$9.0 $\times$ 10$^{59}$ ergs.  Only 25\% of
this ($\sim$2.3$\times$10$^{59}$ ergs)
has gone to heat the gas. 
We have assumed an enthalpy of $4pV$ for each lobe (i.e. $\gamma$=4/3), and
that the lobes can be modeled as cylinders in the plane of the sky \citep{rafferty2006}.
This is the minimum energy of inflation.  If the inflation were in fact supersonic, the
energy imparted to the gas could be considerably larger.
The minimum mean mechanical power of the jet is then
P$_{\rm jet}$ = 6.1 $\times$ 10$^{44}$ ergs~s$^{-1}$,
if we assume the bubbles are buoyantly evolving with a mean speed of 0.5$c_s$ \citep{churazov01}.

We next compute the energy of the outburst assuming that
the shock is caused by an isotropic point explosion.
We model the shock as a one-dimensional point release of energy into a
$\beta$-model atmosphere.  The parameters of the model,
including the energy and age of the burst,
are adjusted to fit the observed surface brightness profile.
This model has been used to constrain the shock parameters for NGC 4636 and other
nuclear outbursts \citep{baldi09}.
First, for the eastern shock, we fit the shock model to the surface
brightness profile, finding that
the shock energy is 4.7 $\times$ 10$^{59}$ ergs and
the shock age is $\sim$3 $\times$ 10$^{7}$ yr.
The latter is better determined, since
it depends largely on the shock radius and its current speed.
Second, for the south-west shock, a similar analysis
yields a shock energy of 7.7 $\times$ 10$^{59}$ ergs and
shock age of $\sim$4 $\times$ 10$^{7}$ yr.
The two ages are fairly similar and this suggests that there is some asymmetry in
the pressure profile.  It is conceivable that the shocks arise from two different outbursts.
If the shocks are part of a front produced
by a single outburst, the total energy would lie (roughly)
between these two values, or $\sim$6.2$\times$10$^{59}$ ergs,
and similarly the shock age would be $\sim$3.5 $\times$ 10$^{7}$ yr.

The mass of the central SMBH can be estimated from the
K-z data for the 3CRR sample \citep{willott03} and is $\sim$4.0 $\times 10^9$ M$_{\odot}$
\citep{mjh07,marconi03}.
Assuming the outburst was powered by the gravitational binding
energy released by accretion, and adopting a mass-energy conversion
efficiency $\epsilon =0.1$ 
and a total outburst energy of between
$\sim$0.9$\times$10$^{60}$~ergs (= $4pV$) and
1.3 $\times$ 10$^{60}$~ergs (= $3pV$ $+$ shock-outburst-energy),
we find that under these assumptions the black hole grew by
$$
\Delta M_{\rm BH} = {(1- \epsilon) \over \epsilon} {E \over c^2} =~\sim10^7~M_\odot.
$$
Here, $\Delta M_{\rm BH}$ accounts for the lost binding energy,
$E$ is the total energy output in mechanical and radiative forms.
We ignore the radiation because it accounts for a negligible
fraction of the current power output.
This growth in mass corresponds to an average growth rate of
$0.3~M_\odot~{\rm yr}^{-1}$ over the past $\sim$$3.5 \times 10^7~{\rm yr}$.
Thus the current outburst is a small ($<$1\%) contribution to
the mass of the central SMBH. 

Can Bondi accretion of the ICM account for the mechanical power of the outburst \citep{allen06}?
It is in principle straightforward
to regulate in the context of feedback models and cooling flows
\citep{nulsen2000,churazov2002,sijacki2007,somerville2008,mcnamara09}
and the X-ray atmosphere provides a steady supply
of fuel.  In relatively low power radio galaxies hosted by
giant ellipticals, Bondi accretion
has been shown to be energetically feasible in the sense that hot
atmospheres probably have a sufficient gas density to supply the
mass required to account for the observed jet powers 
\citep{dimatteo2000,allen06,rafferty2006,mjh09,mcnamara09}.

The average gas density and temperature in the inner 3.2$^{\prime\prime}$
(12 kpc) of 3C\,288's hot halo is n$_e$ = 2.3 $\times$ 10$^{-2}$ cm$^{−3}$ and
2.9$^{-0.2}_{+0.2}$~keV, respectively.  
Using the black hole mass of
$M_{BH}$ = 4.0 $\times$ 10$^9$~M$_\odot$ (see above), 
we find a Bondi accretion rate of
$$
\dot M_{\rm B} = 0.012 \left({n_{\rm e} \over 0.13}\right) \left({kT \over 2.5}\right)^{-3/2} \left({M_{\rm BH} \over 10^9}\right)^2
$$
$$
 = 2.7\times 10^{-2} M_\odot~yr^{-1}.~~~~~~~~~~~~~
$$
This value lies roughly an order of magnitude 
below the $\dot M = 0.3~M_\odot~{\rm yr}^{-1}$
required to power the current outburst
and taken at face value suggests that the current outburst cannot be powered
by Bondi accretion of hot cluster medium.
We caution however that this result relies 
on an extrapolation of the temperature and density profile into the core.
\textit{Chandra}'s resolution is not sufficient to probe the gas on scales of
kiloparsec, and
even a modest increase in density and decrease in temperature of the gas on these spatial
scales could easily balance the Bondi accretion rate with the mechanical power of the outburst.
For example, the gas in the BCGs of hot, non-cool core clusters such as Coma is denser and cooler than that of the
ambient ICM \citep{vikhlinin01b}.
The fact that the shock is detached from the lobes, that no X-ray emission is detectable
from the central AGN, and that the radio galaxy may be transitioning
from FR\,II to FR\,I all support the idea that inflation of the lobes has slowed perhaps due to the energy
supply to the jet being greatly reduced or cut off.
Therefore, it is not surprising that energy released due to the Bondi accretion
is currently far less than the mechanical power of the outburst, if the AGN power
has recently dropped significantly.
Alternatively, it is possible that the current outburst was fueled by accretion of cold gas from,
for example, a dusty disk.  Such accretion
is commonly seen in nearby FR\,IIs such as 3C\,33 \citep{kraft07b}.
There is no evidence of a dusty disk in the \textit{HST} image \citep{dekoff96},
so it is not clear where this cold gas would originate.
Hence, in short, it is plausible for the Bondi accretion to power the current
outburst and cold gas could come from a minor merger, but
the existing data is of insufficient quality to make a definitive statement.

The estimated energy of the shock is roughly twice the value of the minimum
$pV$ work done by the inflation of the lobes, thus demonstrating that supersonic
inflation of the lobes can play a key role in the energy balance of cool core
clusters.  The shock energy is also a significant fraction ($\sim$30\%) of the total
thermal energy of the gas within the radius of the shock.
This possibly suggests that the temperature and thermal energy of the gas in
the core prior to the inflation of the radio lobes was at least 30\% lower.
We conclude that we are most likely witnessing AGN feedback in action, and that the
outburst, which may have been fueled by Bondi accretion of cooling gas at the cluster
center, has likely puffed up the cluster cool core to offset radiative losses.

\subsection{Internal Pressure of the Radio Lobes}

Finally, we use the measured pressure profile of the gas to determine whether the lobes
are at or near equipartition.
To estimate the equipartition magnetic field strength, $B_{eq}$, in the lobes,
we make the conventional assumptions that all relevant features are
cylinders with depths equal to their radius on the plane
of the sky, that the radio spectra are power laws from 10~MHz to 100
GHz, that the filling factor of the emission is unity, and that
equal energies reside in the heavy particles and the relativistic electrons.
With these assumptions, $B_{eq}$ is $\sim$50~$\mu$G over the lobes,
consistent with \citet{bridle89}, and hence,
the equipartition pressure is $\sim$7.7 $\times$ 10$^{-11}$ dyn~cm$^{-2}$.
Using the best-fit models of the $\beta$ profile,
we estimate the thermal gas pressure at the approximate position of the lobes to be
$\sim$1.1 $\times$ 10$^{-10}$ dyn~cm$^{-2}$.
Thus for the case of this FR\,I/FR\,II transitional object, the equipartition pressure of the
lobe is roughly equal to that of the ambient pressure given the uncertainties
of roughly 50\%.  
The result does not change even if we assume that lobes of 3C\,288
do not contain an energetically dominant proton population.
This suggests that this `transitional'
object is, at least in this regard, more similar to the FR\,II radio galaxies than the FR\,Is.
It is typically found that the equipartition pressures of lobes in FR\,I radio galaxies are orders
of magnitude less than the ambient gas, while the $P_{\rm eq}\sim P_{\rm gas}$ for FR\,IIs \citep{croston2005,croston2008}.

\section{Conclusions}

3C\,288 and its gaseous environment provide a laboratory at moderate redshift 
for investigating the interaction between an outburst from a SMBH and the surrounding cluster medium.
Using the 42~ks \textit{Chandra} observations of 3C\,288, we deduce the
following:

\begin{enumerate}
\item We detect two surface brightness discontinuities in the gas at projected
distances of 53.3 (eastern) and 91.1 kpc (southwestern) from the nucleus, which we attribute
to shocks from the supersonic inflation of radio lobes.
\item Under the assumption that the discontinuities are shocks, the gas density jumps
($\rho_{\rm shock}/\rho_0$ $\approx$ 1.48$^{+0.28}_{-0.24}$
and $\rho_{\rm shock}/\rho_0$ $\approx$ 1.75$^{+0.68}_{-0.39}$,
respectively with 90\% uncertainties for the eastern and south-western shocks), yield
shock Mach numbers, 1.33$^{+0.15}_{-0.12}$ and 1.53$^{+0.37}_{-0.21}$,
respectively for the eastern and south-western shocks,
characteristic of a classical shock in a gas with $\gamma$ = 5/3.
The data are not of sufficient quality to detect
the expected jump in temperature at the discontinuity (T$_{\rm shock}$/T$_0$).
\item We measure the energy and age of the shocks to be $\sim$1.6$\times$10$^{60}$ ergs and
3.5$\times$10$^7$ yrs, respectively.
\item The radio lobes are not far from equipartition.
\end{enumerate}

\textit{Chandra} has detected shocks from the supersonic inflation of radio lobes in nearly
two dozen galaxies, groups, and clusters, but the discovery of shocks reported here is
the most distant reported to date.  It is now clear that feedback between cooling gas
and the central SMBH plays a critical role in the evolution of early-type galaxies and the
central regions of groups and clusters.  These outbursts probably suppress star formation
in massive galaxies and are the origin of the exponential decay in the galaxy mass
function at large masses.  A well-selected \textit{Chandra} survey could detect a significant
number of examples of such phenomena in massive systems out to redshift $\sim$0.5.
The lookback time to 3C\,288 is only about 15\% of the Hubble time, so even
this observation has not yet begun to directly study the role of shock-heating in the epoch of
cluster formation ($z$ $\sim$ 1 and beyond).  Given the relative faintness and rarity of even the
most massive clusters beyond $z\sim$0.5, it would be difficult to make a detailed study
with \textit{Chandra} at and beyond the redshift at which clusters are forming.


\acknowledgements

We thank the anonymous referee for suggestions and criticisms which improved
the paper.
DVL thanks R.~Johnson and M.~Machacek for many helpful conversations.
Support for this work was provided by the National Aeronautics and Space
Administration through $Chandra$ Award Number GO8-9111X issued by the $Chandra$
X-ray Observatory Center, which is operated by the Smithsonian Astrophysical
Observatory for and on behalf of the National Aeronautics Space Administration
under contract NAS8-03060.
This research has made use of software provided by the Chandra
X-ray Center in the application packages CIAO and Sherpa.
MJH thanks the Royal Society for a research fellowship.
This research has made use of the NASA/IPAC Extragalactic Database (NED)
which is operated by the Jet Propulsion Laboratory, California Institute of
Technology, under contract with NASA.
This research has made use of NASA's Astrophysics Data System.

%



\clearpage

\end{document}